\documentclass[aps,prl,twocolumn,amsmath,amssymb,superscriptaddress,showpacs,showkeys]{revtex4-2}

\usepackage{dcolumn}
\usepackage{graphicx,color}
\usepackage{multirow}
\usepackage{textcomp}
\usepackage{physics}
\usepackage{dsfont}
\usepackage{relsize}
\usepackage{tikz}	
\usetikzlibrary
{
	arrows,
	calc,
	through,
	shapes.misc,	
	shapes.arrows,
	chains,
	matrix,
	intersections,
	positioning,	
	scopes,
	mindmap,
	shadings,
	shapes,
	backgrounds,
	decorations.text,
	decorations.markings,
	decorations.pathmorphing,	
}
\begin {document}
  \newcommand {\nc} {\newcommand}
  \nc {\beq} {\begin{eqnarray}}
  \nc {\eeq} {\nonumber \end{eqnarray}}
  \nc {\eeqn}[1] {\label {#1} \end{eqnarray}}
  \nc {\eol} {\nonumber \\}
  \nc {\eoln}[1] {\label {#1} \\}
  \nc {\ve} [1] {\mbox{\boldmath $#1$}}
  \nc {\ves} [1] {\mbox{\boldmath ${\scriptstyle #1}$}}
  \nc {\mrm} [1] {\mathrm{#1}}
  \nc {\half} {\mbox{$\frac{1}{2}$}}
  \nc {\thal} {\mbox{$\frac{3}{2}$}}
  \nc {\fial} {\mbox{$\frac{5}{2}$}}
  \nc {\la} {\mbox{$\langle$}}
  \nc {\ra} {\mbox{$\rangle$}}
  \nc {\etal} {\emph{et al.}}
  \nc {\eq} [1] {(\ref{#1})}
  \nc {\Eq} [1] {Eq.~(\ref{#1})}

  \nc {\Refc} [2] {Refs.~\cite[#1]{#2}}
  \nc {\Sec} [1] {Sec.~\ref{#1}}
  \nc {\chap} [1] {Chapter~\ref{#1}}
  \nc {\anx} [1] {Appendix~\ref{#1}}
  \nc {\tbl} [1] {Table~\ref{#1}}
  \nc {\Fig} [1] {Fig.~\ref{#1}}
  \nc {\ex} [1] {$^{#1}$}
  \nc {\Sch} {Schr\"odinger }
  \nc {\flim} [2] {\mathop{\longrightarrow}\limits_{{#1}\rightarrow{#2}}}
  \nc {\textdegr}{$^{\circ}$}
  \nc {\inred} [1]{\textcolor{red}{#1}}
  \nc {\inblue} [1]{\textcolor{blue}{#1}}
  \nc {\IR} [1]{\textcolor{red}{#1}}
  \nc {\IB} [1]{\textcolor{blue}{#1}}
  \nc{\pderiv}[2]{\cfrac{\partial #1}{\partial #2}}
  \nc{\deriv}[2]{\cfrac{d#1}{d#2}}
\title{\textit{Ab initio} prediction of the $^4{\rm He}(d,\gamma)\,^6\rm Li$ big bang radiative capture }
\author{C.~Hebborn}
\email{hebborn@frib.msu.edu}
\affiliation{Facility for Rare Isotope Beams, East Lansing, MI 48824, USA}
\affiliation{Lawrence Livermore National Laboratory, P.O. Box 808, L-414, Livermore, CA 94551, USA}
\author{G.~Hupin}
\affiliation{Université Paris-Saclay, CNRS/IN2P3, IJCLab, 91405 Orsay, France}

\author{K. Kravvaris}
\affiliation{Lawrence Livermore National Laboratory, P.O. Box 808, L-414, Livermore, CA 94551, USA}
\author{S. Quaglioni}
\affiliation{Lawrence Livermore National Laboratory, P.O. Box 808, L-414, Livermore, CA 94551, USA}
\author{P. Navr\'atil}
\affiliation{TRIUMF, 4004 Wesbrook Mall, Vancouver BC, V6T 2A3, Canada}
\author{P. Gysbers}
\affiliation{TRIUMF, 4004 Wesbrook Mall, Vancouver BC, V6T 2A3, Canada}
\affiliation{Department of Physics and Astronomy, University of British Columbia, Vancouver, British Columbia, V6T 1Z1, Canada}
\date{\today}
\begin{abstract}

The rate at which helium ($^4$He) and deuterium ($d$) fuse together to produce lithium-6 ($^6$Li) and a $\gamma$ ray, $^4$He$(d,\gamma)^6$Li, 
is a critical puzzle piece in resolving the 
discrepancy between big bang predictions and astronomical observations for the primordial abundance of $^6$Li. 
The accurate determination of this radiative capture rate requires the quantitative and predictive description of the fusion probability across the big bang energy window ($30$ keV $\lesssim E\lesssim 400$~keV), where measurements are hindered by low counting rates. We present 
first-principles (or, \textit{ab initio}) predictions of the $^4$He$(d,\gamma)^6$Li astrophysical S-factor  
using validated nucleon-nucleon and  three-nucleon interactions derived within the framework of chiral effective field theory. 
By employing the \textit{ab initio} no-core shell model with continuum 
to describe 
$^4{\rm He}$-$d$ scattering 
dynamics and bound $^6\rm Li$ product
on an  equal footing, 
we accurately and consistently determine the contributions of the main electromagnetic transitions driving the radiative capture process.  
Our results  reveal an enhancement of the capture probability below 100 keV owing to previously neglected magnetic dipole (M1) transitions and reduce by an average factor of 7 the uncertainty of the thermonuclear capture rate between $0.002$ and $2$ GK.  
\end{abstract}
\maketitle
%


The isotopes of hydrogen,  helium and lithium present few minutes after the big bang seeded all nucleosynthetic processes responsible for the creation of chemical elements in the Universe.
Although the  big bang nucleosynthesis (BBN)  predictions for the abundances of hydrogen and helium are in agreement with astrophysical observations, they fall short in the cases of lithium isotopes: the abundance of $^7\rm Li$ is overpredicted by a factor of 2-4, and the one of $^6\rm Li$  is underpredicted by {up to} three orders of magnitude
~\cite{F11}. The origin of these discrepancies could be traced to beyond standard model physics or to systematic uncertainties in inferring the primordial abundances from  the composition of metal-poor stars~\cite{Aetal06,CFOY16}. A third possibility is that part of the discrepancy could be explained by inaccuracies in the nuclear reaction rates that are the main inputs to the  BBN reaction network. To arrive at a complete solution of these cosmological lithium problems, it is therefore essential to accurately pin down the astrophysical reaction rates  responsible for  the formation of $^{6,7}\rm Li$ at BBN energies.

The production of $^6\rm Li$ is dominated by the  $^4{\rm He}(d,\gamma)\,^6\rm Li$  radiative capture at BBN energies, from 30~keV to 400 keV, which is poorly known. On the experimental side, there are large discrepancies between existing data sets.  Direct measurements are hindered by the Coulomb repulsion between the $^4{\rm He}$ and $d$ nuclei, that  strongly suppresses the counting statistics. Consequently, there exist only two direct measurements in the BBN energy range, at 94  and  134 keV~\cite{LUNA14}. 
Indirect estimates relating the capture rate with the disintegration of $^6\rm Li$ in the Coulomb field of a heavy target overcome the low statistics but suffer from systematic uncertainties, caused by difficulty to  cleanly separate the nuclear and electromagnetic contributions to the breakup cross section~\cite{BBR86,Ketal91,Hetal10}. Accurate theoretical predictions are  therefore needed to guide the extrapolation of  the existing direct measurements to the whole BBN range of energies. 
{On the theory side, most calculations were carried out in either 
two-body potential models (that neglect the internal structure of the $^4$He and $d$ reactants)~\cite{BZZE90,J93,Metal95,Metal11,Metal16,Getal17} 
or in three-body $^4\rm He$+$p$+$n$ models~\cite{Retal95,Tetal16,BT18}
with an inert $^4\rm He$ core. In both cases, typically the contributions owing to the electromagnetic dipole transitions are approximated.
In the early 2000s, Nollett \etal~\cite{NWS01}  improved these theoretical  predictions by including 
an \textit{ab initio} treatment of all relevant ($^4\rm He$, $d$ and $^6\rm Li$) nuclei, but their analysis still relied on a phenomenological description of the $^4$He-$d$ scattering and suffered from the use of somewhat imprecise variational solutions for the $^4$He and $^6$Li wave functions. Because none of these models 
provides a fully microscopic and consistent description of 
the $^4\rm He$ and $d$ reactants, and of the six-body $^6\rm Li$ bound and 
$^4$He-$d$ 
scattering states, they use phenomenological prescriptions to evaluate the electric dipole (E1) transitions 
and the magnetic dipole (M1) matrix elements are often not computed.} Using these approximations,  quadrupole electric (E2) transitions are predicted to drive the capture above 100~keV,  below which E1 transitions become  dominant. {This work constitutes the first calculations that do not rely on these phenomenological prescriptions and we evaluate the electromagnetic operators exactly.}

In this Letter, 
we present a fully \textit{ab initio} and consistent prediction of the $^4{\rm He}(d,\gamma)\,^6\rm Li$ radiative capture starting from nucleon degrees-of-freedom and their interactions. Scattering and bound states are treated within the same theoretical framework. 
Contrary to previous studies, 
E1 transitions are found to be negligible. 
An enhancement of the capture below 100~keV is instead driven by previously neglected M1 transitions. The uncertainty of the predicted $^4{\rm He}(d,\gamma)\,^6\rm Li$ thermonuclear reaction rates is reduced by an average factor 7 compared to previous evaluations~\cite{NACREII}.

For capture reactions below the Coulomb barrier, the typical observable is the astrophysical S-factor, which is proportional to the cross section~$\sigma$ but is not exponentially suppressed at low energies.
At these energies,
 the capture cross section can be safely approximated by~\cite{D05}
\begin{align}
    \nonumber\sigma(E)&=
    \frac{64\pi^4}{4\pi \epsilon_0 \hbar v}
    \sum_{\kappa\lambda}
    \frac{k_\gamma^{2\lambda+1}}{[(2\lambda+1)!!]^2}\frac{\lambda+1}{\lambda}\\ 
   &\times\sum_{ J_il_i s_i }\frac{\hat J_f^2}{\hat s_P^2 \hat s_T^2 
  \hat l_i^2} \left| \mel{\Psi^{J_f^{\pi_f}T_f}}{\Big|\mathcal{M}^{\kappa \lambda}\Big|}{\Psi^{J_i^{\pi_i}T_i}_{l_is_i}}\right|^2,
  \label{eq2}
\end{align}
where $f$ and $i$ denote respectively the final ($^6$Li) bound-state and initial  ($^4$He-$d$) scattering wavefunction, $P$ and $T$ correspond to the projectile ($d$) and target ($^4\rm He$) nuclei, 
$v$ is the initial relative $P$-$T$ velocity, $\lambda$ is the multipolarity of the electric ($\kappa=\,$E) and magnetic ($\kappa=\,$M) transition operator and the notation $\hat J_f$ stands for $\sqrt{2J_f+1}$.  The quantum numbers $J$, $l$, $s$, $\pi$, and $T$ are respectively the total and orbital angular momenta,  spin, parity, and isospin.  The matrix element in Eq.~\eqref{eq2} is evaluated for E1, E2 and M1 operators, which read
\begin{eqnarray}
    \mathcal{M}^{E\lambda}&=&e\sum_{j=1}^A\frac{1+\tau_{jz}}{2}(\ve r_j-\ve R^{(A)}_{\rm cm})^\lambda\label{eq3}\\
    \mathcal{M}^{M1}&=&\frac{\mu_N}{\hbar c}\sqrt{\frac{3}{4\pi}} \sum_{j=1}^A (g_{lj} L_j+g_{sj}S_j) \label{eq4}
\end{eqnarray}
where $e$ is the electric charge, $\mu_N$ is the nuclear magneton, $\ve R^{(A)}_{\rm cm}$ is the center-of-mass (c.m.) coordinate of the $A$-nucleon system, $g_{sj}$,  $\tau_{jz}$,  $S_j$ and $L_j$ are respectively the gyromagnetic factor, the isospin,   spin and orbital angular momentum (defined with respect to the c.m.) operator of the $j$th nucleon and  $g_{lj}=1$ for proton and $0$ for neutron.  

In the case of $^4{\rm He}(d,\gamma)\,^6\rm Li$,   electric dipole transitions are strongly suppressed because the c.m. of the  $^4\rm He$-$d$ system corresponds to its  center-of-charge~\cite{BT18}.
 Nevertheless,  when the $^4\rm He$ and  $d$ nuclei fuse together to form the bound $^6\rm Li$, this is no longer true  and these E1 transitions can become important.  Models  which do not treat the internal structure of these nuclei  explicitly~\cite{BZZE90,Metal95,Metal11,Metal16,Getal17}  evaluate   E1 transitions  by adopting the experimental masses of the $^4{\rm He}$ and $d$ nuclei, effectively shifting the c.m. away from the center-of-charge and thus generating a small dipole strength.  Recently, the validity of this phenomenological prescription has been questioned since it cannot reproduce the physical energy slope of the S-factor~\cite{BT18}. Moreover, M1 transitions are usually assumed to be negligible, based on the fact that the  operator~\eqref{eq4} can be seen as the sum of a   spin $S_j$  and  total angular momentum $J_j$ contributions, with the second term canceling exactly due to the orthogonality of the initial and final wavefunctions, which are both eigenstates of the underlying microscopic Hamiltonian~\cite{NWS01,BT18,BD83}. Because both E1 and M1 transitions are predicted to be small, the   E2  component typically dominates the capture.   In the present work, we do not rely on these assumptions and compute the transition operators  microscopically starting from the operators~\eqref{eq3}--\eqref{eq4}.

  The no-core shell model with continuum method (NCSMC, see Ref.~\cite{NCSMCPhysScripta} for a recent review) is a tool of choice to predict  $^4{\rm He}(d,\gamma)\,^6\rm Li$ as  it describes accurately both the static properties of light nuclei and their dynamics~\cite{HQN15,DEetal16,PRL11BeNCSMC,HQN19,Ketal20,KQHN20}. The NCSMC $6$-body wavefunction for the $^4$He+$d$ system is given in terms of $^6\rm Li$  no-core shell model (NCSM) wavefunctions $\ket{A\lambda J^\pi T}$ and  continuous $^4\rm He$-$d$ cluster states ${\mathcal{A}}_\nu\ket{\Phi^{J^\pi T}_{\nu r}}$, built from the  $^4\rm He$ and $d$  NCSM states
\begin{eqnarray}
    \ket{\Psi^{J^\pi T}}=\sum_{\lambda}c_\lambda^{J^\pi T}\ket{A\lambda J^\pi T}\hspace{4cm}\nonumber\\
    +\sum_\nu\int_0^{+\infty} dr\, r^2\, \frac{\gamma_\nu^{J^\pi T}(r)}{r}\hat{\mathcal{A}}_\nu\ket{\Phi^{J^\pi T}_{\nu r}}.\hspace{0.5cm}
\end{eqnarray}
The unknown coefficients $c_\lambda^{J^\pi T}$ and $\gamma_\nu^{J^\pi T}$ are obtained by solving the Bloch-\Sch equations, as detailed in Ref.~\cite{NCSMCPhysScripta}. The  E1 matrix elements within the NCSMC formalism are also derived in Ref.~\cite{NCSMCPhysScripta} and expressions for E2 and M1 operators can be obtained in an analogous way, with the exception that they rely on   closure relationships with respect to the NCSM $^6\rm Li$  and the binary $^4\rm He$-$d$  cluster bases, respectively.

Our prediction starts from state-of-the-art  nucleon-nucleon (NN) and three-nucleon (3N) interactions~\cite{vK94,Navratil2007,GQNErr19}  derived from low-energy quantum chromodynamics via chiral effective field theory~\cite{W91}, that  provide an accurate description of both bound and scattering physics. These interactions are  softened  using the similarity renormalization group (SRG) transformation in three-body space with a momentum resolution scale of $\lambda=2$ fm$^{-1}$~\cite{JNF09}.  
The  eigenstates of the aggregate $^6\rm Li$, $^4\rm He$, and $d$ nuclei are obtained using a basis of many-body harmonic oscillator wavefunctions with frequency $\hbar \Omega=20$~MeV and a maximum number $N_{\rm max}=11$ of particle excitation quanta above the lowest energy configuration of the system. Discussions on the  choice of the microscopic Hamiltonian, the  influence of the SRG transformation on the electromagnetic operators and the convergence of our predictions can be found in the Supplemental material (which includes Refs.~\cite{Setal20,FHP12,Getal55,Metal68,Setal14}).

\begin{figure}[h]
	\centering
	{\includegraphics[width=\linewidth]{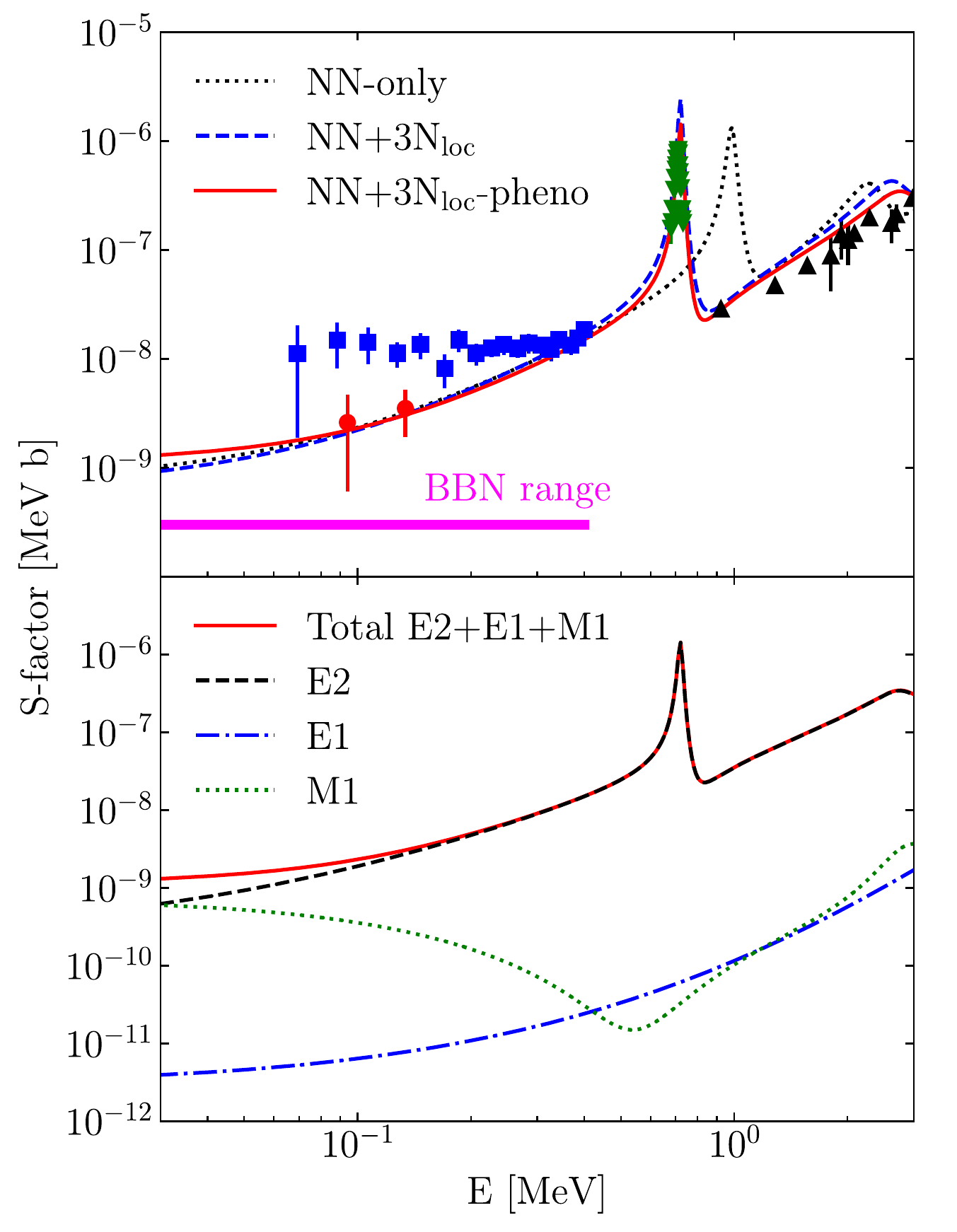}}
	\caption{Top: Predicted S-factor for the $^4{\rm He}(d,\gamma)\,^6\rm Li$ compared with data taken from Refs.~\cite{LUNA14} (red circles), \cite{Ketal91} (blue square), \cite{Metal94}~(green down-triangles) and \cite{Retal81} (black up-triangles). Calculations are  obtained using  the SRG-evolved N3LO NN potential~\cite{EM03} (NN-only) with $\lambda=2$ fm$^{-1}$, the NN+3N$_{\rm loc}$~\cite{vK94,GQNErr19} without (NN+3N$_{\rm loc}$)  and with the phenomenological energy adjustment (NN+3N$_{\rm  loc}$-pheno). Bottom: E2, E1 and M1 components to the predicted S-factor for the $^4{\rm He}(d,\gamma)\,^6\rm Li$ obtained with the NN+3N$_{\rm  loc}$-pheno. }\label{Fig1}
\end{figure}

Our predicted S-factor agrees well with available  existing experimental data~\cite{LUNA14,Ketal91,Metal94,Retal81} (top panel of Fig.~\ref{Fig1}). Overall, when only the SRG-evolved NN potential is considered (NN-only), our calculation reproduces well the magnitude of the data, particularly at low energies where it agrees with the direct measurements of the LUNA collaboration~\cite{LUNA14}. Our results are however incompatible with the ones inferred from breakup data~\cite{Ketal91}, which, as discussed before, have been shown to  suffer from model-dependence~\cite{Hetal10}.  However, this NN-only prediction misses the positions  of the $3^+$ and $2^+$ resonance peaks respectively measured by Mohr \etal~around $E_{3^+}= 0.71$~MeV~\cite{Metal94} and by Robertson \etal~around $E_{2^+}= 2.84$~MeV~\cite{Retal81}. This is unexpected because both the chiral and SRG-induced 3N forces strongly affect the splitting between the $3^+$ and $2^+$ states~\cite{HQN15}.  When  both NN and 3N forces  (both chiral and SRG-induced) are considered, the $^6\rm Li$ $3^+$ and $2^+$  resonances are  in  excellent agreement with the direct measurements of Mohr \etal~and Robertson \etal,  but the ground state (g.s.) is overbound by $\sim$~310~keV (see Supplemental Material). Compared to the NN-only case, the inclusion of the 3N forces modifies  the $^6\rm Li$ g.s. properties, namely its binding energy and asymptotic normalization constants (ANCs) in  the $\ell = 0$  ($\mathcal {C}_0$) and  $\ell = 2$    ($\mathcal {C}_2$) partial-waves in the relative $^4\rm He$-$d$ motion (see Table~\ref{Tab1}),   causing small changes in the magnitude and the slope of the S-factor at low energy~\cite{BB00,Ketal22}. 

\begin{table}
	\begin{tabular}{cr@{\hskip 0.15in}r@{\hskip 0.15in}r@{\hskip 0.15in}r}
		& NN-only &3N$_{\rm loc}$& 3N$_{\rm loc}$-pheno&Exp. or Eval.\\
		\hline \hline
		$E_{\rm g.s.}$ &-1.848&-1.778&-1.474&-1.4743\\
		\multirow{2}{*}{$\mathcal{C}_0$}&\multirow{2}{*}{2.95}&\multirow{2}{*}{2.89}&\multirow{2}{*}{2.62{(4)}}&2.28(7)\\
	&&&&	{2.29(12)}\\
		$\mathcal{C}_2$ & -0.0369&-0.0642&-0.0554{(305)}&-0.077(18)\\
		$\mathcal{C}_2/\mathcal{C}_0$ &-0.013&-0.022&-0.021{(11)}&-0.025(6)(10)\\
		$\mu$ &0.85&0.84&0.84{(1)}&0.8220473(6)\\
		\hline \hline
	\end{tabular}
	\caption{Ground-state properties of $^6\rm Li$ (binding energy $E_{\rm g.s.}$ [MeV], ANCs 	$\mathcal{C}_0$,	$\mathcal{C}_2$ [fm$^{-1/2}$] and magnetic moment $\mu$ [$\mu_N$]) obtained using  the SRG-evolved N3LO NN potential (NN-only) with $\lambda=2$~fm$^{-1}$, the NN+3N$_{\rm  loc}$ without (3N$_{\rm loc}$)  and with the phenomenological energy adjustment (3N$_{\rm loc}$-pheno). The last column lists the experimental  $E_{\rm g.s.}$ and $\mu$~\cite{TILLEY20023}, and ANCs inferred from   a phase shift analysis~\cite{GK99}. The first uncertainty is purely statistical and the second is an estimate of the  systematic error. {The previous evaluation for $\mathcal{C}_0$ of Blokhsintsev \etal~\cite{PhysRevC.48.2390} is also reported (third line).} }\label{Tab1}
\end{table}

  To improve our evaluation  of the S-factor at low energy~\cite{BB00,Ketal22},  we correct   the overbinding of the $^6\rm Li$ g.s.\ by shifting only the  energies of the $1^+$ g.s. and $2^+$ resonant eigenstates of the aggregate $^6$Li system 
  for the full NCSMC to reproduce the experimental energies, as  done in Refs.~\cite{DEetal16,RHNQ16,PRL11BeNCSMC,HQN19}.  This fine-tuning  (NN+3N$_{\rm loc}$-pheno)  impacts mainly the low-energy part of the S-factor and the energy region close to the $2^+$ resonance. This phenomenological correction also brings the  predicted ANCs ($\mathcal{C}_0$ and $\mathcal{C}_2$)  closer to the values inferred from the low-energy  $^6\rm Li$-$^4{\rm He}$ { and $^4\rm He$-$d$  phase shifts in Refs.~\cite{GK99,PhysRevC.48.2390} (last column of Table~\ref{Tab1}).
  The uncertainty associated with our  NN+3N$_{\rm loc}$-pheno results are estimated from  the errors arising from  the truncation of the model space in the number of excitation quanta $N_{\rm max}$ and the choice of the chiral 3N force (see Supplemental material).} {Because our predictions reproduce low-energy capture and elastic-scattering observables (see Supplemental material), the discrepancy between our prediction for $\mathcal{C}_0$ and previous works  extracting  ANCs from  phase shifts is most likely due to systematic uncertainties owing to the use of  optical potentials~\cite{Ketal18,Ketal19,Cetal19} or to the extrapolation procedure  to the experimental binding energy~\cite{Vetal89,PCetal85} that have not been quantified in Refs.~\cite{GK99,PhysRevC.48.2390}. Moreover,  our ratio  $\mathcal {C}_0/\mathcal {C}_2$ is in excellent agreement with the previously extracted evaluation of Ref.~\cite{GK99}, for which systematic  uncertainties have been accounted for.}

The relative importance of the  electromagnetic E2, E1 and M1 transitions  varies with energy (bottom panel of Fig.~\ref{Fig1}). We find that  the E2 transitions dominate the non-resonant and resonant capture, in line with previous works~\cite{BZZE90,J93,Metal95,Metal11,Metal16,Getal17,Retal95,Tetal16,BT18,NWS01}. Different from those studies, we obtain larger E2 strengths, that  can be explained, as the E2 operator~\eqref{eq3} is long-ranged, by the larger amplitude of the $^6\rm Li$ g.s. at large distance, i.e., by
the  larger value of the predicted  ANC  $\mathcal{C}_0$ (second line of Table~\ref{Tab1}). Moreover,  we find a sizeable M1 component that has not been predicted in previous works~\cite{BZZE90,J93,Metal95,Metal11,Metal16,Getal17,Retal95,Tetal16,BT18,NWS01}. This M1 contribution arises from the internal dipole magnetic moments of  the  $^6$Li  and $d$ nuclei, making a full microscopic description essential for an accurate calculation.  The good agreement between our predicted magnetic moment and the experimental one  corroborates our evaluation (last line in Table~\ref{Tab1}).   Finally, our calculations show that the  E1 transitions have a  negligible influence on the S-factor~\footnote{Because only isovector E1 transitions from $T_i=0$ scattering state to the $T_f=1$ component of the $^6\rm Li$ g.s. contribute at low energy~\cite{BT18},  the E1 strength is closely related to the isospin mixing  of  the  $^6\rm Li$ $1^+$ g.s..  In our NCSMC calculation, this mixing is mainly caused by the $T_f=1$ component of the  aggregate $^6$Li g.s., which stays small, i.e., $T_f\leq0.0003$, and therefore  leads to negligible E1 transitions.}, contrary to what it is usually predicted using  phenomenological prescriptions. 

\begin{figure}
    \centering
    \includegraphics[width=\linewidth]{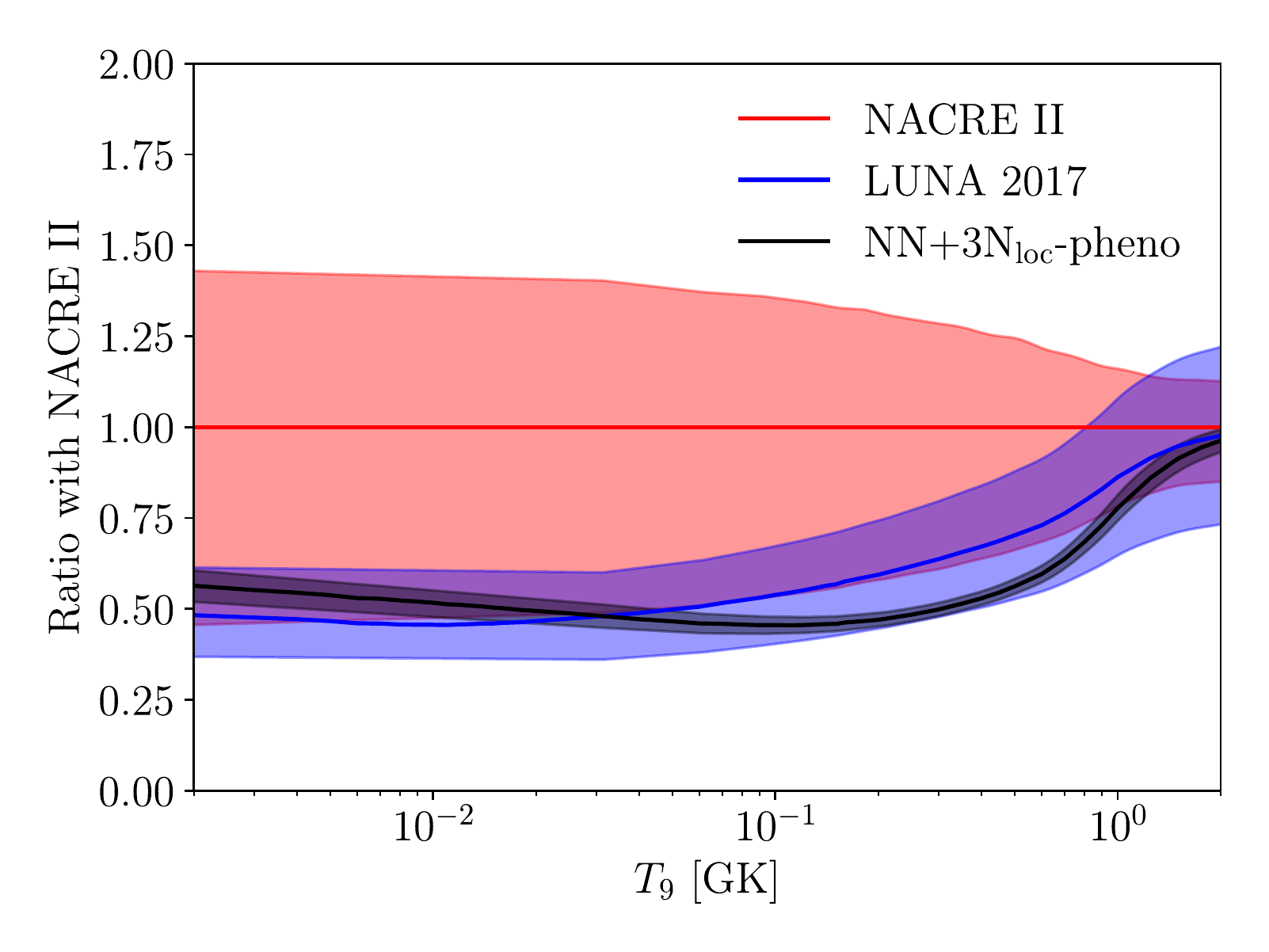}
    \caption{Ratio of the predicted thermonuclear reaction rates (black line) for the $^{4}{\rm He}(d,\gamma)\,^6\rm Li$ with the  NACRE-II evaluation (red line)~\cite{NACREII}  for the $^{4}{\rm He}(d,\gamma)\,^6\rm Li$ for different temperature $T_9$ in GK. Our results are also compared with the recent thermonuclear reaction rate derived from the measurements of the LUNA collaboration (blue line)~\cite{LUNA2017}. The shaded areas correspond to the uncertainty of each calculation (see text for details).
    }
        \label{Fig2}

\end{figure}

From the S-factor at low energy, we obtain a thermonuclear reaction rate  for the $^{4}{\rm He}(d,\gamma)\,^6\rm Li$ (NN+3N$_{\rm loc}$-pheno in Fig.~\ref{Fig2}) with uncertainties reduced by an average factor of 7 compared to the  Nuclear Astrophysics Compilation of REaction rates (NACRE II)~\cite{NACREII}.   Because the low-energy S-factor is dominated by the binding energy and the ANCs of the g.s., the description of which is improved as an effect of the phenomenological correction of the g.s. energy, the uncertainties remain small for all $T_9 \lesssim 2$~GK. Our result is systematically smaller than the NACRE~II rate, but agrees well  with the rates reported by the LUNA collaboration (LUNA 2017)~\cite{LUNA2017}. 
Contrary to our first-principle prediction, both the NACRE~II and LUNA evaluations rely on an extrapolation of experimental data informed by a two-body  $^4\rm He$+$d$ potential  model.

In this Letter, we carried out an \textit{ab initio} prediction  for the  $^4{\rm He}(d,\gamma)\,^6\rm Li$ radiative capture  at BBN energies starting from chiral EFT NN and 3N forces, treating both bound and scattering states within the same formalism and   consistently evaluating  the underlying electromagnetic transitions. In line with previous studies, we find that the E2 transitions dominate the capture at all relevant BBN energies. However, different from the earlier understanding, our results indicate that the M1 transitions become increasingly important at low energies, while the E1 component remains negligible over the whole energy range. The validity of our evaluation is  demonstrated  by the excellent agreement with available S-factor data (both those at low-energy measured by the LUNA collaboration and those in the vicinity of the $3^+$  resonance) and  with the experimental magnetic dipole moment. Our microscopic prediction leads to a systematically lower reaction rate, with an average  reduction of 9\%,  and a factor of 7 smaller uncertainty than the recent NACRE~II evaluation~\cite{NACREII}. 
In this work, we have accounted for systematic uncertainties related to the convergence of our calculations  and the choice of the 3N force. However, we have not accounted for the statistical uncertainties owing to the parameterization of the chiral NN+3N Hamiltonian. We reserve that study for future work.

\begin{acknowledgements}
\textit{Acknowledgments. } C. H. would like to thank D. Baye, D. Phillips and J. Dohet-Eraly for  useful discussions. The work of C. H. is supported by the U.S. Department of Energy, Office of Science, Office of Nuclear Physics, under the FRIB Theory Alliance award no. DE-SC0013617 and under Work Proposal no. SCW0498 under Contract no. DE-AC52-07NA27344. This work was partly supported by LLNL LDRD project 22-LW-003, and also supported by the NSERC Grants No.\ SAPIN-2016-00033, SAPPJ-2019-00039, and PGSD3-535536-2019. TRIUMF receives federal funding via a contribution agreement with the National Research Council of Canada. Computing support for this work came from the Lawrence Livermore National Laboratory (LLNL) Institutional Computing Grand Challenge program. 
\end{acknowledgements}

\bibliographystyle{apsrev}
\bibliography{PRL6Li}

\end{document}


\bce {\Large \bf  Supplemental material for "\textit{Ab initio} prediction of the $^4{\rm He}(d,\gamma)\,^6\rm Li$ big bang radiative capture"}\ece

\section{Microscopic Hamiltonian}

In this study, we  consider two chiral Hamiltonians, both of which accurately predict   the scattering of light nuclei. The N$^3$LO nucleon-nucleon interaction (denoted as NN)~[1]  plus the leading three-nucleon interaction~[2] regulated with either a purely local regulator with a  cutoff of
$\Lambda_{3N}=500$~MeV (denoted as 3N$_{\rm loc}$)~[3], or    using a mix of local and non-local regulators (denoted as 3N$_{\rm lnl}$)~[4].  The  NN+3N$_{\rm loc}$  predicts well the scattering of neutrons on $^4\rm He$    but was found to overbind the $^4\rm He$-$^4\rm He$ system~[5]. Conversely, the use of the non-local regulator (denoted as NN+3N$_{\rm lnl}$) yields a more accurate description of $^4\rm He$-$^4\rm He$ scattering, but degrades somewhat the description of the  $3/2^-$ resonance of  $^5$He~[5,6].   All calculations are performed with  interactions  softened  using the similarity renormalization group (SRG) transformation in three-body space with a momentum resolution scale of $\lambda=2$ fm$^{-1}$.  We include the 3N force matrix elements (chiral and SRG-induced) up to a total number of single-particle quanta for the three-body basis of  $E_{3\rm max}=14$.   The   relative motion betwwen $^4\rm He$ and $d$ is expanded on a harmonic oscillator basis, with a maximum number of quanta $N_{\rm rel}=11$. The 3N  contributions are computed  up to $N_{\rm rel}=9$. Our results are not impacted by  these truncations as  we obtain exactly the same S-factor using a stricter truncation  $N_{\rm rel}=7$.

Because the NCSMC and the aggregate NCSM  ground-state (g.s.) energies are expected to agree at convergence, we   determine which chiral Hamiltonian more accurately describes  the  $^4{\rm He}(d,\gamma)\,^6\rm Li$ radiative capture by analyzing the NCSM $^6$Li, $^4$He and $d$  g.s. energies.
Table~\ref{Tab1} shows these 
g.s. energies for various model space sizes $N_{\rm tot}=N_0+N_{\rm max}$, with $N_0$  the number of quanta in the  lowest energy configuration and with $N_{\rm max}=4$ up to  $N_{\rm max}=12$   quanta of excitation. 
The infinite $N_{\rm max}$  g.s. energy values for  $^6\rm Li$  and $^4\rm He$ nuclei  are obtained by extrapolation with an exponential convergence ansatz~[7]
\begin{equation}
E(N_{\rm max})= E_{\infty}+a \exp \left[-b N_{\rm  max}\right].
\end{equation}	 
The extrapolated $^6\rm Li$ g.s. energies\footnote{We note that the  NCSM $^6\rm Li$  g.s.\  energy obtained with NN+3N$_{\rm loc}$  differs slightly from the one published in Table~I of Ref. [8]. This difference is caused first by the fact that we use the  updated version of 3N$_{\rm loc}$~[3] and second  by a different choice of model space. In the NCSMC description of the $^4{\rm He}(d,\gamma)\,^6\rm Li$ S-factor, the two components (discrete aggregate and continuous microscopic cluster) of the models space are characterized by different $N_0$ values: the aggregate $^6\rm Li$ states  have $N_0=2$ since there are two nucleons in the $p$ shell in the lowest energy configuration, while the binary cluster states   have $N_0=0$ because  the nucleons  in the lowest energy configurations are all in the $s$ shell. In Ref. [8], the authors  chose to  keep the same model space sizes $N_{\rm tot}$ for both  aggregate and binary cluster states (hence different $N_{\rm max}$ values). Here we choose to combine states  with the same $N_{\rm max}$ values (hence different $N_{\rm tot}$). The choice adopted in this work is more consistent as we allow the same number of quanta of excitation in the description of  $^4 \rm He$, $d$, $^6\rm Li$ and the $^4 \rm He$-$d$ relative motion. The  $ E(12)$ value given in  Ref.~[8] should be compared with our results at $N_{\rm max}=10$.}, obtained using   NN+3N$_{\rm loc}$ and NN+3N$_{\rm lnl}$ interactions, do not exactly reproduce  the experimental value:  NN+3N$_{\rm loc}$  overbinds $^6\rm Li$ by $ 250$~keV, while NN+3N$_{\rm lnl}$  underbinds it by $590$~keV. We observe similar features for  $^4\rm He$, with the binding energy overestimated by $100$~keV for NN+3N$_{\rm loc}$ and underestimated by 40~keV  for NN+3N$_{\rm lnl}$.  At convergence,  the $^6\rm Li$ binding energy compared to the $^4\rm He$-$d$ threshold will therefore be overpredicted by 150~keV with NN+3N$_{\rm loc}$ and underpredicted by 550~keV with  NN+3N$_{\rm lnl}$. Because NN+3N$_{\rm loc}$   more closely reproduces the experimental relative binding energy, this interaction  leads to a more accurate S-factor for $^4{\rm He}(d,\gamma)\,^6\rm Li$ and is the one that we adopt for our \textit{ab initio} prediction.

\begin{table}[h]
	
	\centering
	
	\begin{tabular}{cccccc}
		\hline\hline 
		&\multicolumn{2}{c}{$^6\rm Li$}&\multicolumn{2}{c}{$^4\rm He$}&\multicolumn{1}{c}{$d$}\\
		$E_{\rm g.s.}$ [MeV]&NN+3N$_{\rm loc}$& NN+3N$_{\rm lnl}$&NN+3N$_{\rm loc}$& NN+3N$_{\rm lnl}$&NN\\\hline
		$  	E(4)$&-26.7117&-25.7637&-26.5539&-26.5923&-1.2939 \\
		$  	E(6)$&-29.5032&-28.5640&-27.6899 &-27.6831&-1.9199 \\
		$  	E(8)$&-30.8845&-29.9581&-28.1393&-28.1124&-1.9633 \\
		$  	E(10)$&-31.5636&-30.6656& -28.3145&-28.2753&-2.1172 \\
		$  	E(12)$&-31.9098&-31.0420& -28.3505&-28.3072&-2.1352\\     
		$E_{\infty}$&-32.24(1)&-31.40(2)&-28.40(2)&-28.36(2)&\\
		Exp.&\multicolumn{2}{c}{-31.99}&\multicolumn{2}{c}{-28.30}&\multicolumn{1}{c}{-2.22}\\
		\hline\hline
	\end{tabular}
	
	\caption{NCSM ground-state  energies ($E_{\rm g.s.}$) of $^6\rm Li$, $^4\rm He$ and $d$ nuclei obtained from the NCSM using the chiral Hamiltonians  NN+3N$_{\rm loc}$ and NN+3N$_{\rm lnl}$ (see text for details) compared to experiment. 	The error estimate includes only the uncertainties due to the extrapolation.}\label{Tab1}
\end{table} 
\pagebreak
\section{Impact of the 3N force on observables} 
\begin{figure}[h!]
	\centering
	{}{\includegraphics[clip,trim=2cm 2.2cm 2cm 2.2cm,width=0.6\linewidth]{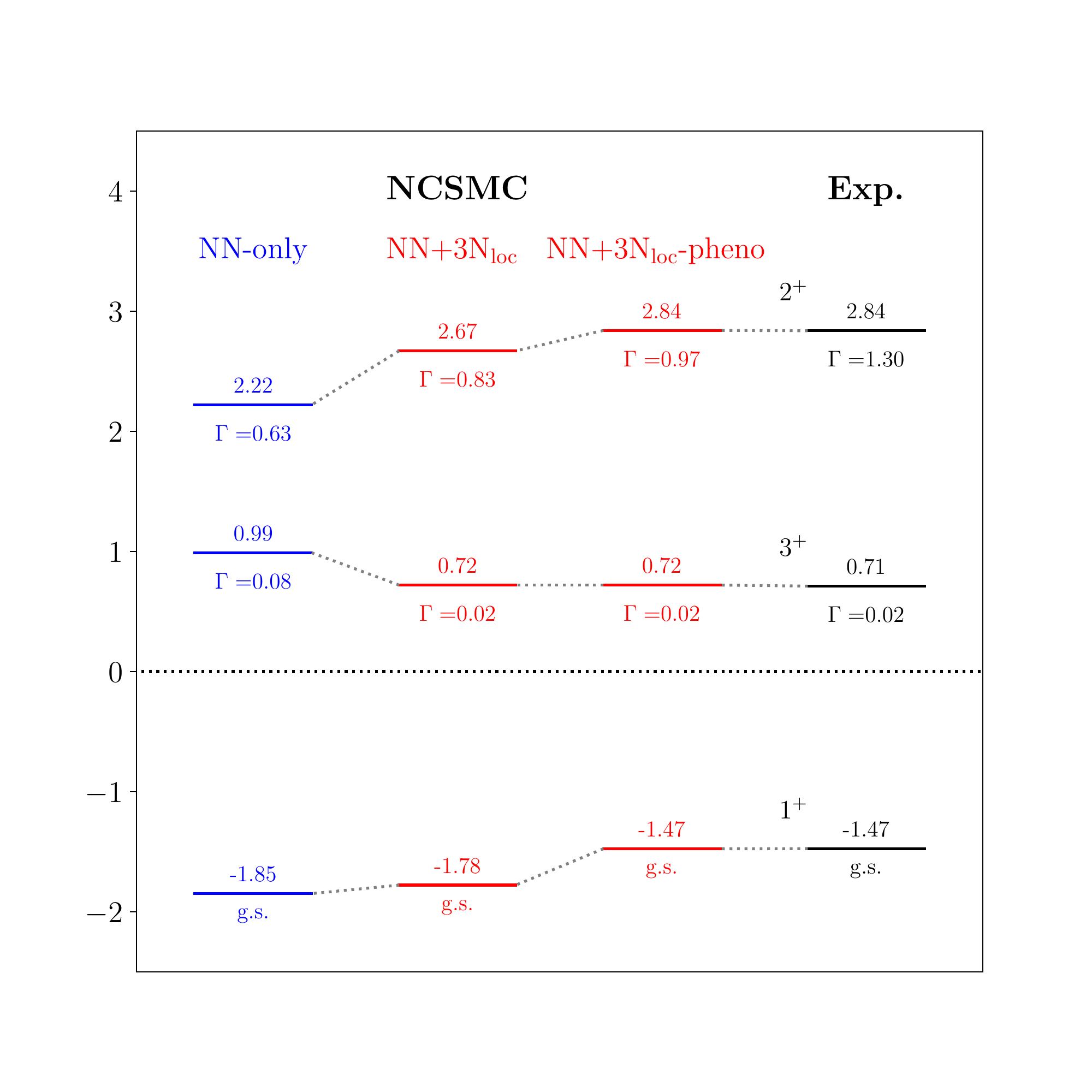}}
	
	\caption{Low-lying spectrum of $^6\rm Li$ obtained  with the NCSMC at  $N_{\rm max}=11$ using the SRG-evolved N$^3$LO NN potential~[1] (NN-only) with $\lambda=2$ fm$^{-1}$, the NN+3N$_{\rm loc}$ [2,3] without (NN+3N$_{\rm loc}$)  and with the phenomenological energy adjustment (NN+3N$_{\rm loc}$-pheno) compared to experiments~[9]. The zero energy is taken as the predicted (resp. experimental) $^4 \rm He$-$d$ threshold for NCSMC (resp. Exp.). } \label{Fig1a}
\end{figure}

The impact of the chiral  3N$_{\rm loc}$ and SRG-induced 3N forces  on the $^6\rm Li$ low-lying spectrum  predicted by the NCSMC is illustrated in Fig.~\ref{Fig1a}.  The zero of energy is taken  for the NCSMC calculations as  the predicted $^4 \rm He$-$d$ threshold, obtained from the  binding energies of the NCSM $^4 \rm He$ and $d$ nuclei. Compared to the NN-only calculations, including the 3N force (SRG-induced and chiral) significantly improves  the $3^+$-$2^+$ splitting and   the position of the $3^+$ resonance, which now has the correct energy and width. Even though the inclusion of the 3N$_{\rm loc}$ force  slightly ameliorates  the overbinding of  $^6\rm Li$, the ground-state energy is still over-predicted by 310~keV.   This can be partly explained by the  slower convergence rate of the $^4 \rm He$-$d$  threshold and aggregate $^6\rm Li$  ground-state [see $E(10)$ in Table~\ref{Tab1}]. Nevertheless, even at convergence, i.e., $N_{\rm max}\to \infty$, the $^6\rm Li$ ground state is expected to remain overbound by approximately 150~keV, due to the present choice of Hamiltonian. As mentioned in the Letter, we correct for this overbinding by adjusting the NCSM eigenenergy of the  aggregate $^6\rm Li$   ground state in such a way that the NCSMC calculation reproduces the experimental binding energy (NN+3N$_{\rm loc}$-pheno in Fig.~\ref{Fig1a}).

\begin{figure}[h!]
	\centering
{\includegraphics[clip,trim=0.5cm 0.5cm 0.5cm 0.5cm,width=0.6\linewidth]{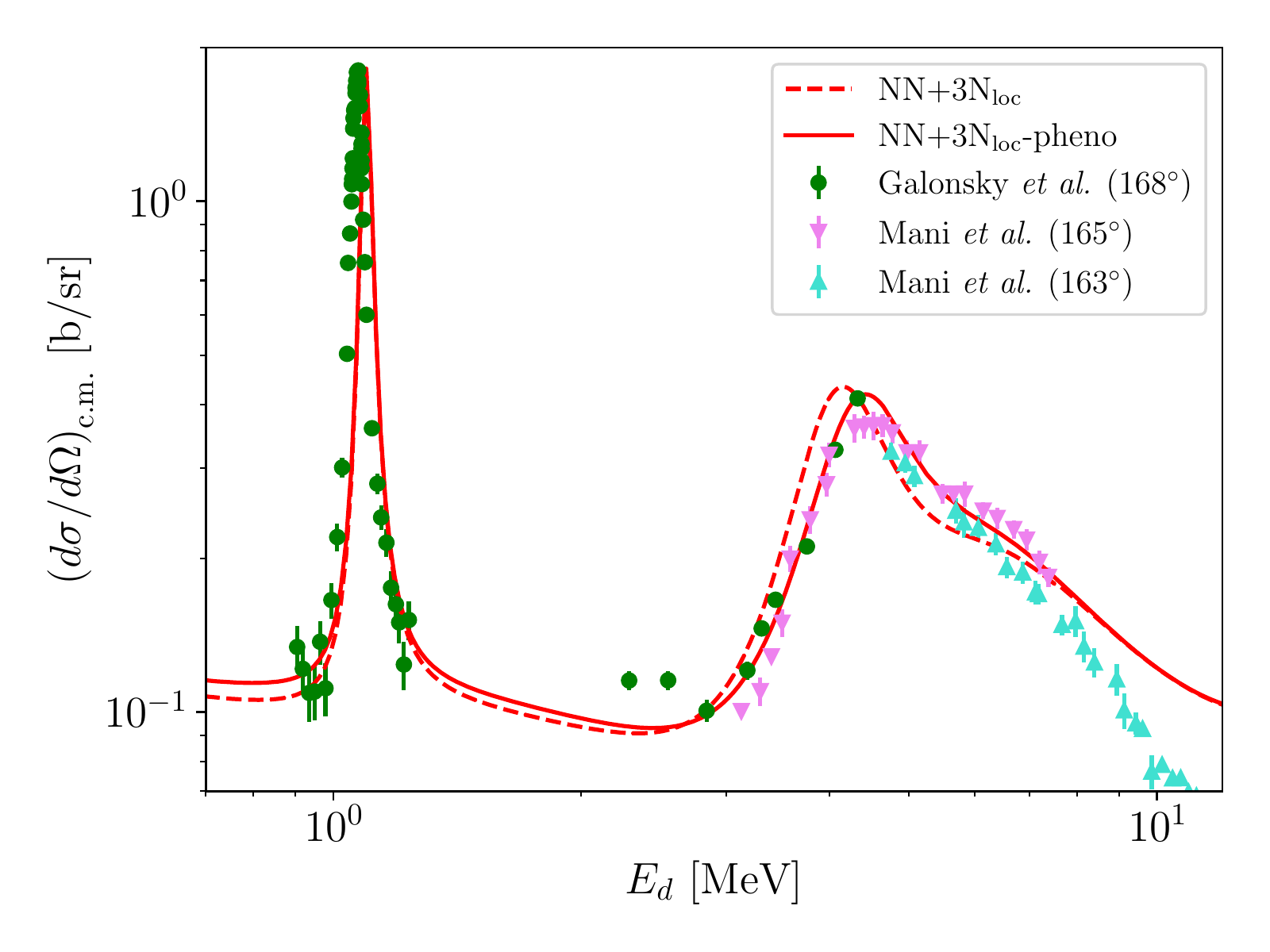}}
	\caption{Both NCSMC predictions obtained without (NN+3N$_{\rm loc}$)  and with the phenomenological adjustement (NN+3N$_{\rm loc}$-pheno) reproduce the experimental elastic-scattering $^4{\rm He}(d,d)\,^4\rm He$ cross section at  the deuteron backscattered angle $\theta_d=164^\circ$~[10,11].} \label{Fig1b}
\end{figure}

Our NCSMC predictions without any phenomenological adjustment also  agrees  with  elastic-scattering data for  $^4 {\rm He}(d,d)\,^4\rm He$~[9,10] (Fig.~\ref{Fig1b}).   In particular, the  NCSMC calculations reproduce  the $3^+$  resonance peak  at 1.065 MeV.  Including the phenomenological correction (NN+3N$_{\rm loc}$-pheno)  modifies   slightly the non-resonant part   at low energy and improves the description of the  $2^+$ resonance peak at 4.26~MeV.

\pagebreak
\section {Convergence of the  $S$-factor for $^4{\rm He}(d,\gamma)\,^6\rm Li$}

\begin{figure}[h!]
	\centering
	\includegraphics[width=0.6\linewidth]{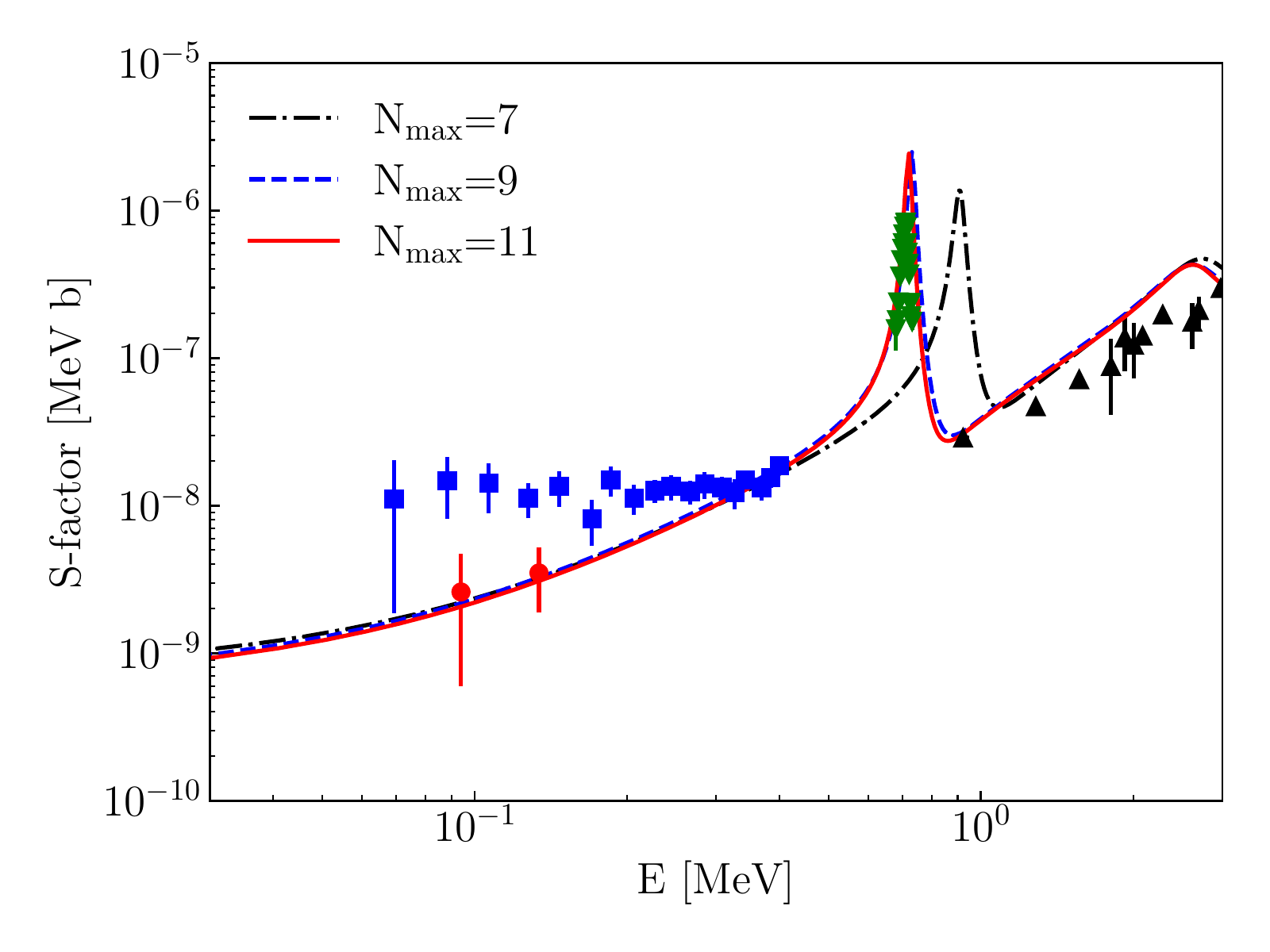}
	\caption{Convergence of the  S-factor for $^4{\rm He}(d,\gamma)\,^6\rm Li$ obtained using the Hamiltonian  NN+3N$_{\rm loc}$ with $N_{\rm max}$. The experimental data are from Refs. [12] (red circles), [13] (blue square), [14]~(green down-triangles) and [15] (black up-triangles).
	}\label{Fig2}
\end{figure}

We illustrate in Fig.~\ref{Fig2}  the convergence of the   S-factor for  $^4{\rm He}(d,\gamma)\,^6\rm Li$   with respect to the HO model-space size $N_{\rm max}$. Our calculations converge rapidly;  already at $N_{\rm max}=7$ the  non-resonant part of the S-factor is accurate. Allowing for larger model spaces ($N_{\rm max}\geq 9$)   improves the S-factor in the vicinity of the $3^+$ resonance peak, which now  falls on top of the data, but leaves the low-energy part largely unchanged.  For the convergence in the number of deuteron pseudostates, we observe a similar  pattern to the one displayed in Fig. 1 of Ref. [8].
\pagebreak

\section {Negligible influence of SRG evolution on  M1 contributions}
 
\begin{figure}[h]
	\centering
	\includegraphics[width=0.6\linewidth]{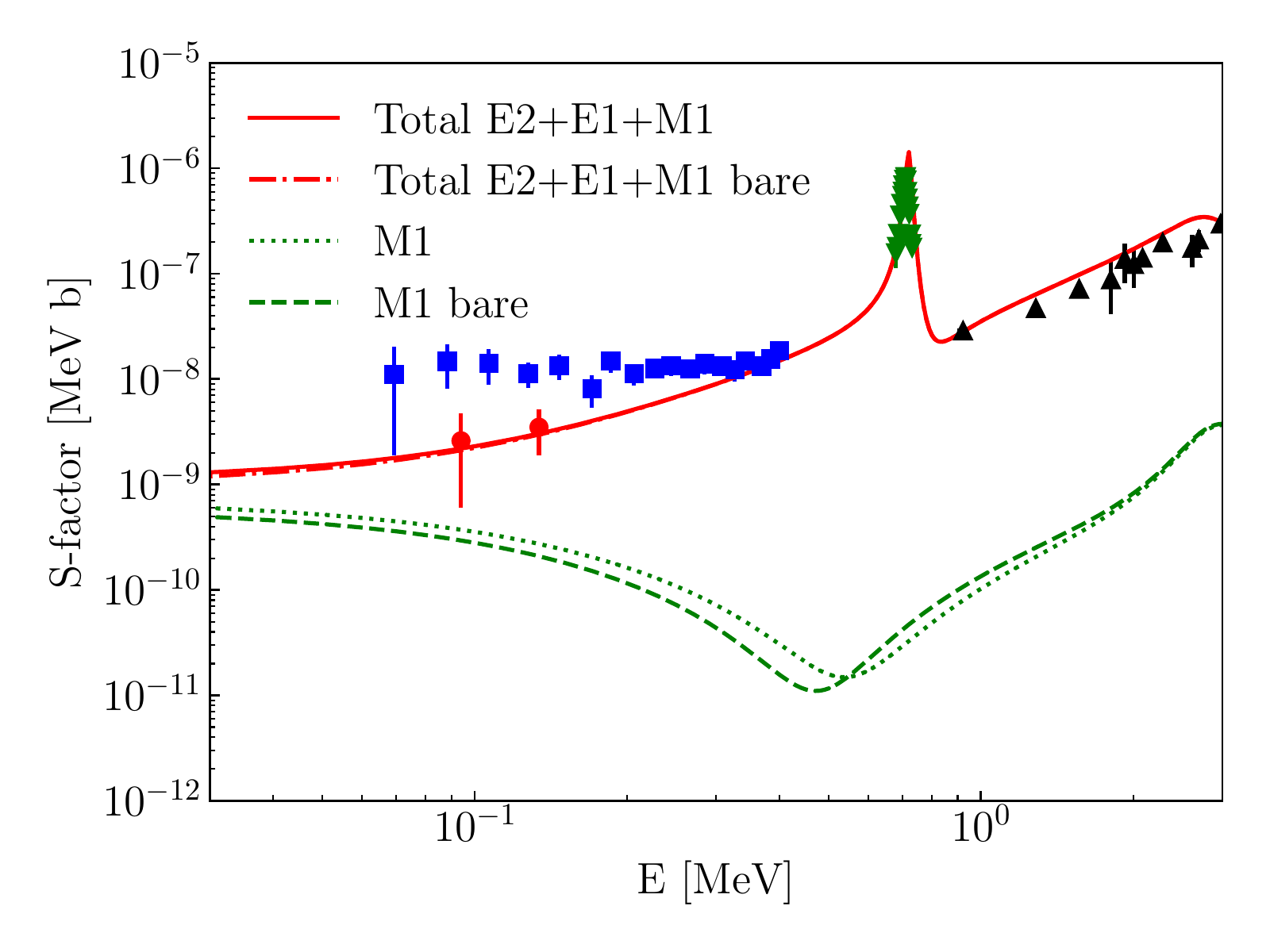}
	\caption{Influence of the SRG transformation on the M1 operator. The experimental data follow the same legend as in Fig.~\ref{Fig2}.
	} \label{Fig3}
\end{figure}

 We  also investigate the importance of the SRG transformation on the electromagnetic  matrix elements. Because E1 and E2 operators are long-range, their matrix elements depend mainly  on the  wavefunctions at large distances and  the SRG transformation has  a negligible impact~[16]. For the M1 operator, the situation is  different  as it  is the sum of an angular momentum and spin operators, therefore being more sensitive to the  short-range physics. In the NCSMC description of $^4{\rm He}(d,\gamma)\,^6\rm Li$,  the electromagnetic strengths can be decomposed in four components, i.e.,  the  matrix element of the aggregate $^6\rm Li$ states, the  matrix elements between aggregate  $^6\rm Li$ state and binary  $^4\rm He$-$d$ cluster basis states (and vice versa), and the  matrix element between binary $^4\rm He$-$d$ cluster states~(see Eq.~(82) from Ref.~[17]).  The analysis of each contribution separately reveals that the M1 strengths result  mainly  from the internal M1 transitions of $^6\rm Li$ and $d$  NCSM ground states. Interestingly, the minimum in the M1 contribution around 0.5~MeV and its enhancement at low energies are caused by interferences between these internal  transitions, indicating that this shape will only  be obtained within models that include the structure  of the $^6\rm Li$ and $d$ nuclei. To estimate the effect  of the  neglected SRG evolution of the M1 operator, we compare  in Fig.~\ref{Fig3} calculations obtained with these dominant   matrix elements   evaluated with   the M1 operator SRG-evolved  in two-body space (M1) and with a  bare M1 operator (M1 bare).  The modifications are small  and have a negligible effect on the total S-factor (Total E2+E1+M1 compared with Total E2+E1+M1 bare).  

\section {Thermonuclear reaction rates}
We give in Table~\ref{TabReactionRate} the  thermonuclear reaction rates for the  $^4{\rm He}(d,\gamma)\,^6\rm Li$ at different values of the temperature $T_9$. The total uncertainty $\epsilon_{\rm tot}$ is obtained from the uncertainties $\epsilon_{N_{\rm max}}$ due to the truncation in $N_{\rm max}$  and $\epsilon_{\rm 3N}$ due to the specific choice of   3N force
\begin{equation}
    \epsilon_{\rm tot}=\sqrt{\epsilon_{N_{\rm max}}^2+\epsilon_{\rm 3N}^2}.
\end{equation}
The first, $\epsilon_{N_{\rm max}}$, is estimated from the difference between the rates obtained with NCSMC calculations obtained at $N_{\rm max}=9$ and at $N_{\rm max}=11$, both using the phenomenological adjustments of the $1^+$ and $2^+$ energies. The 3N force uncertainty $\epsilon_{\rm 3N}$ is  evaluated  from the difference between the NCSMC predictions using NN+3N$_{\rm loc}$ and NN+3N$_{\rm lnl}$,   also obtained with the phenomenological correction. In the range of  temperature $T_9$ considered here, the uncertainty is dominated by the choice of the 3N force.

\begin{table}[]
    \centering
    \begin{tabular}{ccccc}
    \hline \hline 
    $T_9$ [GK]& $N_A \langle\sigma v\rangle$ [cm$^3$ mol$^{-1}$ $s^{-1}$] &$\epsilon_{tot}$ [cm$^3$ mol$^{-1}$ $s^{-1}$]&$\epsilon_{N_{\rm max}}$ [cm$^3$ mol$^{-1}$ $s^{-1}$]&$\epsilon_{\rm 3N}$ [cm$^3$ mol$^{-1}$ $s^{-1}$]\\ \hline 
        0.002 &1.679$\,10^{-23}$& 1.317$\,10^{-24}$& 6.317$\,10^{-25}$& 1.155$\,10^{-24}$\\
        0.003& 2.212$\,10^{-20}$& 1.721$\,10^{-21}$ &8.231$\,10^{-22}$& 1.512$\,10^{-21}$\\
        0.004& 2.031$\,10^{-18}$& 1.568$\,10^{-19}$& 7.476$\,10^{-20}$& 1.378$\,10^{-19}$\\
        0.005& 5.021$\,10^{-17}$& 3.851$\,10^{-18}$& 1.830$\,10^{-18}$& 3.388$\,10^{-18}$\\
        0.006& 5.774$\,10^{-16}$& 4.400$\,10^{-17}$& 2.085$\,10^{-17}$ &3.875$\,10^{-17}$\\
        0.007& 4.048$\,10^{-15}$ &3.065$\,10^{-16}$ &1.448$\,10^{-16}$& 2.702$\,10^{-16}$\\
        0.008& 2.014$\,10^{-14}$& 1.515$\,10^{-15}$& 7.134$\,10^{-16}$& 1.337$\,10^{-15}$\\
        0.009& 7.802$\,10^{-14}$& 5.836$\,10^{-15}$ &2.739$\,10^{-15}$& 5.153$\,10^{-15}$\\
        0.010& 2.502$\,10^{-13}$& 1.860$\,10^{-14}$& 8.702$\,10^{-15}$ &1.644$\,10^{-14}$\\
        0.011& 6.921$\,10^{-13}$& 5.118$\,10^{-14}$& 2.386$\,10^{-14}$& 4.527$\,10^{-14}$\\
        0.012 &1.702$\,10^{-12}$& 1.252$\,10^{-13}$& 5.816$\,10^{-14}$& 1.108$\,10^{-13}$\\
        0.013& 3.803$\,10^{-12}$&2.781$\,10^{-13}$& 1.288$\,10^{-13}$& 2.465$\,10^{-13}$\\
        0.014& 7.850$\,10^{-12}$& 5.708$\,10^{-13}$& 2.635$\,10^{-13}$& 5.064$\,10^{-13}$\\
        0.015& 1.515$\,10^{-11}$& 1.096$\,10^{-12}$& 5.042$\,10^{-13}$& 9.733$\,10^{-13}$\\
        0.016& 2.765$\,10^{-11}$& 1.989$\,10^{-12}$& 9.116$\,10^{-13}$& 1.768$\,10^{-12}$\\
        0.018& 8.002$\,10^{-11}$& 5.697$\,10^{-12}$& 2.592$\,10^{-12}$& 5.073$\,10^{-12}$\\
        0.020& 1.997$\,10^{-10}$& 1.407$\,10^{-11}$& 6.355$\,10^{-12}$& 1.255$\,10^{-11}$\\
        0.025& 1.243$\,10^{-9}$& 8.548$\,10^{-11}$& 3.786$\,10^{-11}$& 7.663$\,10^{-11}$\\
        0.030& 5.006$\,10^{-9}$& 3.359$\,10^{-10}$& 1.457$\,10^{-10}$& 3.027$\,10^{-10}$\\
        0.040 &3.801$\,10^{-8}$& 2.436$\,10^{-9}$& 1.009$\,10^{-9}$& 2.217$\,10^{-9}$\\
        0.050& 1.606$\,10^{-7}$& 9.857$\,10^{-9}$ &3.877$\,10^{-9}$& 9.062$\,10^{-9}$\\
        0.060& 4.829$\,10^{-7}$& 2.843$\,10^{-8}$ &1.055$\,10^{-8}$& 2.640$\,10^{-8}$\\
        0.070& 1.165$\,10^{-6}$& 6.593$\,10^{-8}$& 2.291$\,10^{-8}$& 6.183$\,10^{-8}$\\
        0.080 &2.416$\,10^{-6}$& 1.316$\,10^{-7}$& 4.248$\,10^{-8}$& 1.246$\,10^{-7}$\\
        0.090& 4.483$\,10^{-6}$& 2.357$\,10^{-7}$& 7.003$\,10^{-8}$& 2.250$\,10^{-7}$\\
        0.100 &7.651$\,10^{-6}$& 3.889$\,10^{-7}$& 1.053$\,10^{-7}$& 3.744$\,10^{-7}$\\
        0.110& 1.224$\,10^{-5}$& 6.025$\,10^{-7}$& 1.468$\,10^{-7}$& 5.844$\,10^{-7}$\\
        0.120 &1.858$\,10^{-5}$& 8.878$\,10^{-7}$& 1.918$\,10^{-7}$& 8.668$\,10^{-7}$\\
        0.130 &2.703$\,10^{-5}$& 1.256$\,10^{-6}$& 2.362$\,10^{-7}$& 1.234$\,10^{-6}$\\
        0.140 &3.798$\,10^{-5}$& 1.720$\,10^{-6}$& 2.749$\,10^{-7}$& 1.697$\,10^{-6}$\\
        0.150 &5.181$\,10^{-5}$& 2.289$\,10^{-6}$& 3.012$\,10^{-7}$& 2.270$\,10^{-6}$\\
        0.160& 6.893$\,10^{-5}$& 2.978$\,10^{-6}$& 3.076$\,10^{-7}$& 2.962$\,10^{-6}$\\
        0.180 &1.147$\,10^{-4}$& 4.760$\,10^{-6}$& 2.254$\,10^{-7}$ &4.755$\,10^{-6}$\\
        0.200& 1.785$\,10^{-4}$& 7.169$\,10^{-6}$& 5.048$\,10^{-8}$& 7.169$\,10^{-6}$\\
        0.250 &4.404$\,10^{-4}$& 1.665$\,10^{-5}$& 2.204$\,10^{-6}$& 1.651$\,10^{-5}$\\
        0.300 &8.920$\,10^{-4}$& 3.262$\,10^{-5}$& 7.779$\,10^{-6}$& 3.168$\,10^{-5}$\\ 
        0.350& 1.589$\,10^{-3}$& 5.719$\,10^{-5}$& 1.860$\,10^{-5}$& 5.408$\,10^{-5}$\\
        0.400& 2.586$\,10^{-3}$& 9.281$\,10^{-5}$& 3.667$\,10^{-5}$ &8.526$\,10^{-5}$\\
        0.450 &3.944$\,10^{-3}$& 1.424$\,10^{-4}$& 6.430$\,10^{-5}$& 1.271$\,10^{-4}$\\
        0.500 &5.730$\,10^{-3}$& 2.100$\,10^{-4}$& 1.045$\,10^{-4}$& 1.822$\,10^{-4}$\\
        0.600& 1.095$\,10^{-2}$& 4.234$\,10^{-4}$& 2.404$\,10^{-4}$& 3.485$\,10^{-4}$\\
        0.700 &1.937$\,10^{-2}$& 8.026$\,10^{-4}$& 4.943$\,10^{-4}$ &6.324$\,10^{-4}$\\
        0.800 &3.269$\,10^{-2}$& 1.439$\,10^{-3}$& 9.315$\,10^{-4}$& 1.097$\,10^{-3}$\\
        0.900& 5.302$\,10^{-2}$& 2.417$\,10^{-3}$& 1.611$\,10^{-3}$& 1.802$\,10^{-3}$\\
        1.000 &8.239$\,10^{-2}$ &3.783$\,10^{-3}$& 2.567$\,10^{-3}$& 2.779$\,10^{-3}$\\
        1.250& 2.031$\,10^{-1}$ &8.817$\,10^{-3}$& 6.109$\,10^{-3}$& 6.358$\,10^{-3}$\\
        1.500& 3.908$\,10^{-1}$& 1.541$\,10^{-2}$& 1.077$\,10^{-2}$& 1.103$\,10^{-2}$\\
        1.750 &6.292$\,10^{-1}$& 2.247$\,10^{-2}$& 1.576$\,10^{-2}$ &1.602$\,10^{-2}$\\
        2.000 &8.973$\,10^{-1}$& 2.919$\,10^{-2}$& 2.053$\,10^{-2}$& 2.075$\,10^{-2}$ \\ \hline \hline
    \end{tabular}
    \caption{Predicted thermonuclear reaction rate  for the $^4{\rm He}(d,\gamma)\,^6\rm Li$ at different temperature $T_9$. The total error estimate $\epsilon_{\rm tot}$ is obtained from the uncertainty resulting from the $N_{max}$ truncation $\epsilon_{N_{\rm max}}$ and the choice of the 3N force $\epsilon_{\rm 3N}$. }
    \label{TabReactionRate} 
\end{table}
\clearpage
\noindent

\textbf{References}\\
{[1] D. R. Entem and R. Machleidt, Phys. Rev. C \textbf{68}, 041001(R) (2003).}\\
{[2] U. van Kolck, Phys. Rev. C \textbf{49}, 2932 (1994).}\\
{[3] D. Gazit, S. Quaglioni, and P. Navr\'atil, Phys. Rev. Lett. \textbf{122}, 029901(E) (2019).}\\
{[4] V. Som\'a, P. Navr\'atil, F. Raimondi, C. Barbieri, and T.  Duguet, Phys. Rev. C \textbf{101}, 014318 (2020).}\\
{[5] K. Kravvaris, S. Quaglioni, G. Hupin, P. Navr\'atil, arXiv:2012.00228 (2020).}\\
{[6] K. Kravvaris, K. R. Quinlan, S. Quaglioni, K. A. Wendt, and P. Navr\'atil, Phys. Rev. C \textbf{102}, 024616 (2020).}\\
{[7] R. J. Furnstahl, G. Hagen, and T. Papenbrock
Phys. Rev. C \textbf{86}, 031301(R) (2012).}\\
{[8] G. Hupin, S. Quaglioni, and P. Navr\'atil, Phys. Rev. Lett. \textbf{114}, 212502  (2015).}\\
{[9] D. Tilley, C. Cheves, J. Godwin, G. Hale, H. Hofmann, J. Kelley, C. Sheu, and H. Weller, Nucl. Phys. A \textbf{708}, 3 (2002).}\\
{[10] A. Galonsky \etal Phys. Rev. \textbf{98}, 586 (1955).}\\
{[11]  G. S. Mani and A. Tarratts, Nucl. Phys. A \textbf{107}, 624	(1968).}\\
{[12] M. Anders \etal  (LUNA Collaboration), Phys. Rev. Lett. \textbf{113}, 042501 (2014).}\\
{[13] J. Kiener \etal  Phys. Rev. C \textbf{44}, 2195 (1991).}\\
{[14] P. Mohr \etal Phys. Rev. C \textbf{50}, 1543 (1994).}\\
{[15] R. G. H. Robertson \etal  Phys. Rev. Lett. \textbf{47}, 1867 (1981).}\\
{[16] M. D. Schuster, S. Quaglioni, C. W. Johnson, E. D. Jurgenson, and P. Navr\'atil
	Phys. Rev. C \textbf{90}, 011301(R) (2014).}\\
	{[17] P. Navr\'atil, S. Quaglioni, G. Hupin, C. Romero-Redondo,
and A. Calci, Phys. Scr. \textbf{91}, 053002 (2016).}